\documentclass[pra,laps,10pt,twocolumn,floatfix,showpacs]{revtex4-1}
\usepackage{amsmath,graphicx,amssymb,braket,xcolor,subfigure,upgreek,bbm}

\newcommand{\identity}{\mathbbm{1}}

\begin{document}

\title{Subradiance and entanglement in atoms with several independent decay channels}

\author{Martin Hebenstreit, Barbara Kraus, Laurin Ostermann and Helmut Ritsch }
\email{Laurin.Ostermann@uibk.ac.at}
\affiliation{Institut f\"ur Theoretische Physik, Universit\"at
Innsbruck, Technikerstrasse 25, A-6020 Innsbruck, Austria}

\date{\today}

\begin{abstract}
Spontaneous emission of a two--level atom in free space is modified by other atoms in its vicinity leading to super- and subradiance. In particular, for atomic distances closer than the transition wavelength the maximally entangled antisymmetric superposition state of two individually excited atomic dipole moments possesses no total dipole moment and will not decay spontaneously at all. Such a two-atom dark state does not exist, if the atoms feature additional decay channels towards other lower energy states. However, we show here that for any atomic state with $N-1$ independent spontaneous decay channels one can always find a $N$-particle highly entangled state, which completely decouples from the free radiation field and does not decay. Moreover, we show that this state is the unique state orthogonal to the subspace spanned by the lower energy states with this property. Its subradiant behavior largely survives also at finite atomic distances.
\end{abstract}


\maketitle

The decay of an excited atomic state towards lower lying states via spontaneous emission is one of the most striking consequences of the quantum nature of the free radiation field~\cite{dirac1927quantum}. Heuristically introduced even before e.g. by Einstein, the spontaneous emission rate $\Gamma = \frac{\omega^3 \mu^2}{3\epsilon_0 \pi  \hbar c^3 }$, called A-coefficient, is proportional to the squared transition dipole moment $\mu^2$ between the upper and lower atomic state and the third power of the transition frequency $\omega$~\cite{weisskopf1935probleme}. 

Interestingly, it turns out that for several particles the emission process is not independent but can be collectively enhanced or reduced depending on the atomic arrangement~\cite{lehmberg1970radiation}. It was already noted some time ago that these superradiant and subradiant collective states, where a single excitation is distributed over many particles, are entangled atomic states~\cite{ficek2002entangled,bellomo2016quantum}. Although a recent classical coupled dipole model also leads to subradiance-like phenomena~\cite{facchinetti2016storing}, the most superradiant and the perfect dark states for two two-level atoms with states $(\ket{g},\ket{e})$ are the maximally entangled symmetric and antisymmetric dipole moment superposition states
\begin{equation}
\ket{\psi_{\pm}} = (\ket{eg}\pm\ket{ge})/\sqrt{2}.
\label{psidark}
\end{equation}
  
\begin{figure}[h]
 \includegraphics[width=0.9\columnwidth]{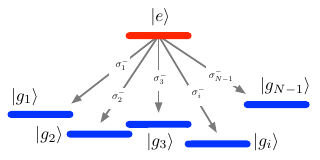}
 \caption{\emph{Level scheme of an atom with several independent decay channels of different polarization or frequency}.}
 \label{fig1}
\end{figure}

While superradiance on a chosen transition persists in the case, when the atom possesses more than one decay channel, there is no completely dark state for two multilevel atoms with several decay channels from a single excited state $\ket{e}$ to several  lower lying states $\ket{g_i}$ as schematically depicted in Fig.~\ref{fig1}. Hence, in practise the observation of subradiant states is much more difficult than seeing superradiance, as all other decay channels need to be excluded~\cite{guerin2016subradiance}.

In this paper, we introduce a new class of subradiant or dark states appearing for atoms with several independent transitions. As a key result of this work we find that for systems of $N$ particles one can construct highly multi-partite entangled states, where all $N-1$ independent decay channels are suppressed. For these states the total dipole moments on all of these $N-1$ transitions simultaneously vanish and at least in principle the optical excitation in this state will be stored indefinitely. 

After introducing our model atom system and the generalized, unique multi-atom dark states, we will discuss their special entanglement properties and possible quantum information processing based schemes to prepare them. In the final part of the paper we numerically study some more realistic geometries using $Lambda-$type atoms, where the effect of dipole-dipole coupling leads to a reduced but still finite lifetime as population of the single excited dark state via decay from multiply excited states will accumulate.

Interestingly, a related phenomenon appears in V-type atoms with two excited and one ground state, where a single ground state atom can prevent the decay of several excitations. This is only shortly exhibited in Appendix II to avoid confusion with the $\Lambda$-geometry. 

\textbf{\emph {Model:}} Let us assume a collection of $N$ identical $N$-level emitters with a set of $N-1$ low energy eigenstates $\left \vert g_i\right\rangle$, where $i \in \lbrace 1, \ldots, N \rbrace$, which are dipole coupled to a higher energy state $\left\vert e \right\rangle$ separated by the excitation energies $\hbar \omega_i$ (see Fig.~\ref{fig1}). The atomic center of mass motion is treated classically with fixed positions $\mathbf{r}_i$ for $i \in \lbrace 1, \ldots, N \rbrace$, which we will assume to be close-by, i.e. within a cubic wavelength. For each atom $i$ and transition $j$ we define individual Pauli ladder operators $ \sigma_j^{i \,\pm}$ describing transitions between the $i$-th atom's excited state $\left\vert e \right\rangle_i$ and each of its $j$ lower energy states $\ket{g_j}_i$, respectively.

The coupling of each atomic transition $i,j$ to the electromagnetic vacuum leads to an individual free space decay rate $\Gamma^i_j$. As all atoms are coupled to the same vacuum modes, these decay rates are modified by pairwise interactions with neighboring atomic transitions $k,l$, which upon elimination of the field modes can be described by mutual decay rates $\Gamma ^{ik}_{jl}$, with $\Gamma ^{ii}_j =\Gamma_j$~\cite{lehmberg1970radiation, ostermann2012cascaded}. Note, that in addition to modifying the decay properties, the collective coupling of the atoms to the vacuum modes also induces energy level shifts $\Omega^{ik}_{jl}$ as presented in~\cite{ostermann2012cascaded}. For simplicity we will first assume a highly symmetric arrangement of the particles, so that all particles acquire equal energy shifts $\Omega^{ik}_{jj}=\Omega_j$, which can be incorporated in effective transition frequencies~\cite{kramer2015generalized}. Our central interest targets the modifications of the emission rates via the collective decay mechanism.  

 In terms of the operators defined above with the excited state energy set to zero, the dipole coupled atomic Hamiltonian is given by~\cite{kramer2015generalized}
\begin{equation}
\label{H_dipole}
H = \sum_{i, j} {-\bar \omega^i_j} \, \sigma ^{i \, -}_j \sigma^{i \, +}_j +  \sum_{i \neq k} \sum_j \Omega^{ik}_j \, \sigma^{i \, +}_j \sigma^{k \, -}_j.
\end{equation}

The full dynamics of the coupled open system including decay is then governed by a master equation for the density matrix $\rho$ of the whole system of $N$ multilevel emitters, reading
\begin{equation}
 \frac{\partial \rho} {\partial t} = i [\rho ,H]+\mathcal{L}[\rho ].
 \label{master}
\end{equation}

Following standard quantum optical assumptions and methods, the effective Liouvillian for the collective decay summed over all transitions and atom pairs reads~\cite{freedhoff1979collective,kramer2015generalized}
\begin{equation}
\begin{aligned}
\mathcal{L}[\rho ] =& \frac{1}{2} \sum_{i, k} \sum_j \Gamma^{ik}_j \left[ 2 \sigma^{i \, -}_j \rho \, \sigma^{k \, +}_j \right. \\
  - & \left. \sigma^{i \, +}_j \sigma^{k \, -}_j \rho -\rho \, \sigma^{i \, +}_j \sigma^{k \, -}_j \right].
\end{aligned}
\end{equation}

While this can be a rather complex and complicated expression for a general atomic arrangement~\cite{kramer2015generalized}, in the case of atoms much closer to each other than the transition wavelength, all $\Gamma^{ik}_j = \Gamma_j$ become approximately independent of the atomic indexes ($i, k$), essentially reducing to a single constant $\Gamma_j$. For simplicity, we will first also assume equal decay rates on all transitions $\Gamma_j=\Gamma$, i.e. equal dipole moments and Clebsch-Gordon coefficients. This will hardly be exactly true for any real atomic configuration (besides a $J=0$ to $J=1$ transition), but it will not change the essential conclusions below.

\textbf{\emph {Collective atomic dark states:}} 
Obviously any atomic density matrix $\rho_g$ involving only ground states, $\ket{g_i}$, occupations trivially will be stationary under $\mathcal{L}$ with $L \left[ \rho_g \right] = 0$. The much more interesting task is to find states, $\rho_e$, featuring atomic excitations, which will not decay under $\mathcal{L}$ and are stationary with eigenvalue zero, i.e. $L \left[ \rho_e \right] = 0$.

For the case of two two-level atoms  such dark states are well known and have been confirmed experimentally decades ago~\cite{grangier1987quantum}. A generic dark state can then be written as in Eq.~(\ref{psidarkn}), i.e., $\ket{\psi_\text{d}^2}=\ket{\psi_{-}}$. As a central claim of this work we show that this formula can be generalized to the case of $N$ atoms with $N-1$ independent optical transitions between the upper state $\ket{0} = \ket{s_0}=\ket{e}$ and $N-1$ lower states $\ket{i} = \ket{s_i}=\ket{g_i} $ in the form
\begin{equation}
\ket{\psi_\text{d}^N} =  \frac{1}{\sqrt{N!}} \sum_{\pi \in S_N} \mbox{sgn}(\pi) \bigotimes_{i} \ket{s_{\pi(i)}} ,
\label{psidarkn}
\end{equation}
where the sum runs over all permutations $\pi$ of $N$ elements. Using the criterion for pure states to be stationary under $\mathcal{L}$ given in~\cite{kraus2008stationaryconditions}, we show in Appendix~\ref{Appendix_I} that this $N$ $N$-level state of total spin $0$ is the unique stationary state which is orthogonal to the subspace where all particles are in $\ket{g_i}$ for some $i$. A symmetric variant of this state, denoted by $\ket{\psi_{\text{sr}}^N}$, with all positive signs will be its super-radiant analogue.

\begin{figure}[h]
 \includegraphics[width=0.9\columnwidth]{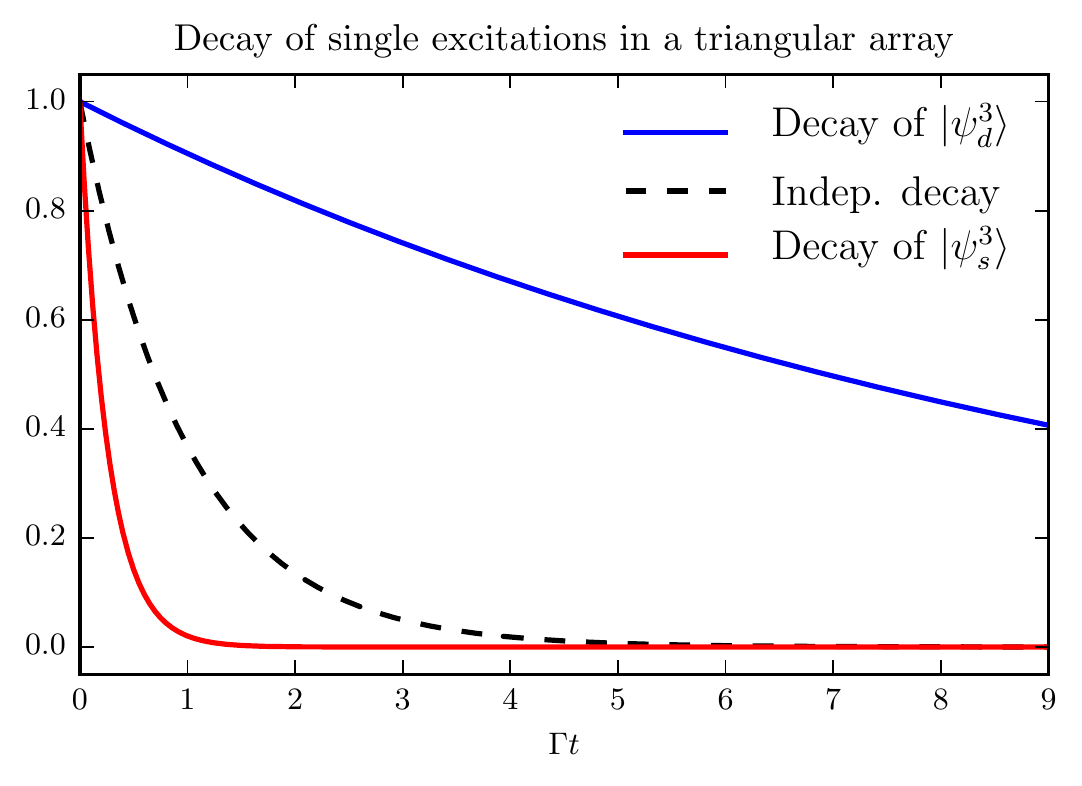}
 \caption{\emph{Upper state population decay of three close by interacting $\Lambda$-type atoms in a triangular configuration of size $d \ll \lambda$ with $\Gamma^{12}_j= 0.95 \Gamma$ starting from the ideal dark state  (blue line). For comparison the dashed black line gives the case of independent atom decay, while the red line corresponds to a fully symmetric state with a superradiant decay on both transitions.}}
 \label{fig2}
\end{figure}

For a 3-level $\Lambda$-atom ($N=3$) one explicitly gets
\begin{equation}
\begin{aligned}
\ket{\psi_\text{d}^3} &=&\frac{1}{\sqrt{6}} \{ \ket{eg_1g_2}+\ket{g_1g_2e}+\ket{g_2eg_1} \\
&& - \ket{eg_2g_1}-\ket{g_2g_1e}-\ket{g_1eg_2} \} .
\label{psidark3}
\end{aligned}
\end{equation}
which is a state within the set of maximally entangled tripartite states of qutrits~\cite{devicente2013mes}.

This state has zero total dipole moment $\mu_j= \left \langle\sum_i \sigma^i_j \right \rangle =0$ on both transitions. As shown in Fig.~\ref{fig2}, for a sub-wavelength triangular atomic configuration it exhibits subradiant decay subject to the master equation Eq.~(\ref{master}). As the atoms will not only undergo collective decay but also experience energy shifts from the resonant dipole-dipole coupling $\Omega^{ik}_j$ in Eq.~\eqref{H_dipole}, the dark states in Eq.~\eqref{psidarkn} in general are not eigenstates of $H$. Hence, dynamic mixing with other states induces a finite lifetime as it is the case for conventional dark states~\cite{plankensteiner2015selective}. Note that the subradiant states discussed here are not the dark states appearing in two laser excitation of $\Lambda$-type systems. There, a particular superposition of ground states decouples from the laser excitation for each atom separately~\cite{dalton1985coherent} and leads to a coherent population trapping without any excitation.

As an important consequence of the uniqueness of the dark state, no such state can exist for a smaller number of atoms. Hence, when considering $M$ atoms with $N-1$ independent optical transitions between the upper state $\ket{e}$ and their $N-1$ lower states $\ket{g_i}$, where the emitted photons on each transition are distinguishable, the following picture emerges: for $M < N$ only ground states are stationary under $\mathcal{L}$. In case $M>N$, however, extra stationary states involving excitations can be found. They are given by tensor products of states that are stationary for parts of the system and superpositions of these states. To give a simple example for the case $M=6$ and $N=3$ the states
\begin{equation}
\ket{\psi} = (\alpha \ket{\psi_\text{d}^3} \otimes \ket{\psi_\text{d}^3} + \beta \ket{g_i g_j} \otimes \ket{\psi_\text{d}^3} \otimes \ket{g_k})/\sqrt{2}
\end{equation}
are dark for any $\alpha, \beta \in \mathbb{C}$.

\textbf{\emph {Entanglement properties of the dark states}}: 
The dark states $\ket{\psi_\text{d}^N}$ given in Eq.~(\ref{psidarkn}) are complex entangled states, whose mathematical properties have been considered before and will be shortly recapitulated here. For instance, it has been shown that $\ket{\psi_\text{d}^N}$ can be used to solve the Byzantine agreement problem, the $N$ strangers problem, the secret sharing problem, and the liar detection problem~\cite{fitzi2001byzantine, cabello2002strangers}. Moreover, it has been shown that there exists no local hidden variable model describing quantum predictions for the state $\ket{\psi_\text{d}^N}$. To this end, generalized Bell inequalities that are violated by $\ket{\psi_\text{d}^N}$ have been constructed for any $N$~\cite{cabello2002strangers}. 

What makes these states so useful for the above mentioned tasks are their rather special entanglement properties~\cite{cabello2002strangers}, which we will discuss in the following. First, note that the bipartite entanglement shared between any of the particles and the rest of the ensemble is maximal. In other words, the reduced density matrix for any particle $\rho_j$, which is obtained by performing the partial trace of $\ket{\psi_\text{d}^N}$ over all particles but particle $j$, is proportional to identity. This also implies that the state is contained in the maximally entangled set~\cite{devicente2013mes} as it cannot be reached from any other (not  Local Unitary (LU)-equivalent \footnote{Note that LUs do not alter the entanglement properties in the state as they can be applied reversibly. Hence, only non-LU transformations are relevant in this context.}) state by Local Operations and Classical Communication (LOCC). Furthermore, it can be shown easily that these states can be transformed via LOCC into another pure, entangled state.

An important property of $\ket{\psi_\text{d}^N}$ is that for all invertible operators $S$, $\ket{\psi_\text{d}^N} \propto S^{\otimes N} \ket{\psi_\text{d}^N}$, which implies several important properties. First, if one particle is measured in any basis and the measurement outcome as well as the chosen basis are announced, then the other $N-1$ particles can be transformed deterministically to the state $\ket{\psi_\text{d}^{N-1}}$ by performing LUs only. Note, however, that $\ket{\psi_\text{d}^{N-1}}$ is not dark for $N-1$ atoms with $N-1$ decay channels.

Second, the geometric measure of entanglement~\cite{barnum2001geometric} can be computed easily and one obtains that $E_g(\ket{\psi_\text{d}^N}) =1- \max_{\ket{a_1}, \ldots, \ket{a_N}} \left|\braket{a_1, \ldots, a_N|\psi_\text{d}^N} \right|^2 = 1- \frac{1}{N!}$ (see Appendix~\ref{Appendix_I})~\cite{hayashi2008geometric}. Due to that, $W = \frac{1}{N!} \identity - \ket{\psi_\text{d}^N}\bra{\psi_\text{d}^N}$ is an entanglement witness (see e.g.~\cite{terhal2001winesses,*lewenstein2000witnesses}), i.e. $\operatorname{tr}(W \rho_\text{sep}) \geq 0$ for any separable state $\rho_\text{sep}$, and there exists a state $\rho$ s.t. $\operatorname{tr}(W \rho) < 0$.

Using this witness, it can be shown, that the entanglement of the states $\ket{\psi_\text{d}^N}$ is persistent under particle loss. In other words, if the partial trace of the state $\ket{\psi_\text{d}^N}$ over any particle is performed, the resulting state is still entangled. For $N=3$ this can be verified by calculating the negativity of the reduced density matrix. For higher $N$, however, the statement can be proven using the witness mentioned before and even holds under loss of a large portion of the particles.

\textbf{\emph {Preparing collective dark states}}: 
As these totally antisymmetric states have intricate entanglement properties they cannot be prepared from a product ground state with simple local operations.  In the following we propose two ways to prepare the state $\ket{\psi_\text{d}^3}$, which can be generalized to prepare $\ket{\psi_\text{d}^N}$ for $N>3$.

In both methods we initially prepare the state $\ket{\psi_-} = (\ket{01}-\ket{10})/\sqrt{2}$ for two of the particles denoted as particles 1 and 2, which can be achieved by applying a $CNOT$ to the two particles in the initial product state $(\ket{0} -\ket{1})/\sqrt{2} \otimes \ket{1}$, respectively.

In the first method we then prepare particle 3 in the state $\ket{2}$ and apply the 3-qutrit gate $e^{- i 2\pi/9 (X \otimes X \otimes X + h.c.)}$, where $X=\ket{1}\bra{0}+\ket{2}\bra{1}+\ket{0}\bra{2}$, in order to obtain the state $\ket{\psi_\text{d}^3}$ up to local phase gates. This preparation procedure can be easily verified noting that $X^3 = \identity$.

Alternatively, we prepare particle 3 in $\ket{+} = (\ket{0}+\ket{1}+\ket{2})/\sqrt{3}$ and apply the two-qutrit unitary $U = \ket{0}\bra{0}  \otimes X + \ket{1}\bra{1} \otimes X^2 + \ket{2}\bra{2} \otimes \identity$ on the particle pairs $(3,1)$ and (3,2) in order to obtain $\ket{\psi_\text{d}^3}$. 

Let us point out, that given $\ket{\psi_\text{d}^{N-1}}$ for $N-1$ particles, the state $\ket{\psi_\text{d}^{N}}$ can be prepared by preparing particle $N$ in $1/\sqrt{N} \sum_{i=0}^{N-1} (-1)^{(N-1)(1+i)} \ket{i}$ and applying $U = \sum_{i=0}^{N-1} \ket{i}\bra{i} \otimes X^{i + 1}$, where $X=\ket{0}\bra{N-1} + \sum_{i=0}^{N-2} \ket{i+1}\bra{i}$, to all particle pairs $(N,j)$. Hence, $\ket{\psi_\text{d}^{N}}$ can be prepared recursively. In a similar manner the state $\ket{\psi_\text{sr}^N}$ can be prepared. This can be achieved by using $\ket{\psi_+}$ instead of $\ket{\psi_-}$ as the initial state of particles 1 and 2 and omitting the minus sign in the initial state of particle $N$. However, the properties of $\ket{\psi_\text{sr}^N}$ are very different from $\ket{\psi_{d}^N}$, as e.g. $\ket{\psi_\text{sr}^N}$ has much less symmetries. Another difference can be found in the geometric measure of entanglement $E_g(\ket{\psi_\text{sr}^N}) = 1 - N!/N^N$~\cite{hayashi2008geometric,aulbach2010geometricqubits}, which is smaller than $E_g$ for the dark state.

\textbf{\emph {Dissipative dynamics of dipole-dipole coupled atom arrays:}}
In the following we will exhibit the collective dynamics of the system for various configurations, where we restrict ourselves to the case of three atoms with two decay channels, i.e. three $\Lambda$-systems, in a equilateral triangle or, alternatively, an equidistant chain. As shown in Fig.~\ref{fig2} at a small but finite distance the dark state will decay much slower than independent atoms. In addition, changing all the minus signs to plus signs leads to superradiance on both transitions.
\begin{figure}[h]
 \includegraphics[width=0.9\columnwidth]{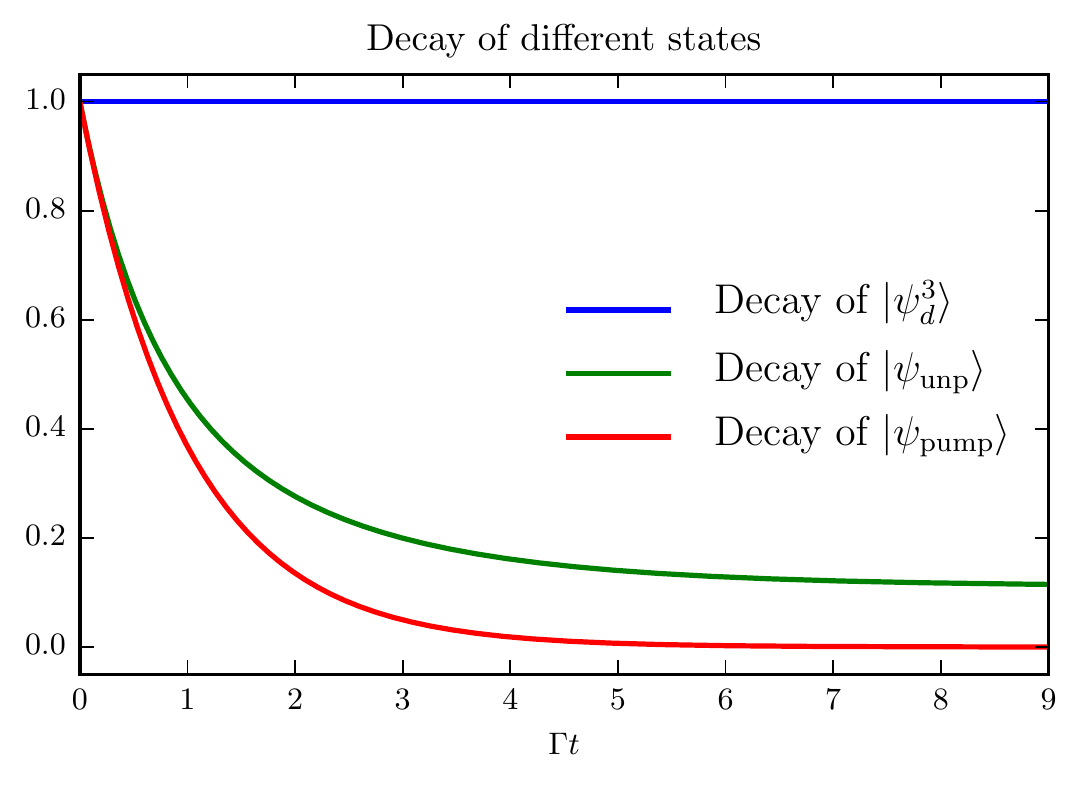}
 \caption{ \emph{ Upper state population decay for three close spaced lambda-type atoms for three different single excited initial  states. The blue line corresponds to the pure dark state $\ket{\psi_\text{d}^3}$, the red line gives the case of the two atom dark state for optically pumped atoms $ 1/\sqrt{2}(\ket{eg_1} - \ket{g_1e}) \ket{g_1} $ and green corresponds to an unpolarized product state $ 1/\sqrt{2}(\ket{e g_1}-\ket{g_1e})\ket{g_2} $ involving all three atomic states }}
 \label{fig3}
\end{figure}

As a second example we demonstrate the surprising fact that a nearby atom in the final state of the transition can be used to suppress the decay of an atom excitation. For this we start from an anti-symmetric two-atom state $\ket{\psi^2_\text{d} }= (\ket{g_1,e}-\ket{e,g_1})/\sqrt{2}$, which will not decay on the first transition to $\ket{g_1}$. This state, however, decays on the second transition towards $ (\ket{g_1,g_2}-\ket{g_2,g_1})/\sqrt{2}$. Now, let us add a third atom in either of the two ground states. As shown in Fig.~\ref{fig3}, a third atom prepared in $\ket{g_1}$ will not prevent decay (red line), while a third nearby atom in the state $\ket{g_2}$ partially prevents total decay and results in a finite excited state population probability at long times(green line). Hence, after some time the system has either decayed to  $(\ket{g_1,g_2,g_2}-\ket{g_2,g_1,g_2})/\sqrt{2}$ or ends up in the dark state $\ket{\psi_\text{d}^3}$. Thus preparing two atom dark states in the vicinity of an unpolarized state of independent ground state atoms provides a method for probabilistic preparation of the dark states.

It is known that in spatially extended systems with non-uniform radiative coupling coefficients $\Gamma^{ik}$, there is no perfect dark state but only long lived subradiant states~\cite{zoubi2008bright}. In this case the free space spontaneous decay from a multiply excited state can also sometimes end up in such a gray state, thus creating entanglement by collective emission~\cite{ostermann2012cascaded}, which has been proposed and used for tailored deterministic entanglement generation between the ground states of interacting $\Lambda$-atoms~\cite{gonzalez2015deterministic, svidzinsky2015quantum}.

Recently we also became aware of another, even more complicated scheme based on three $M$-shaped 5-level atoms interacting via three coupled cavities and lasers~\cite{shao2016dissipative}. Here a dark tripatite entangled state involving only atomic ground states of the three atoms with very similar properties to our case $N=3$ is found to be efficiently populated by a deterministic dissipative preparation scheme.
\begin{figure}
 \includegraphics[width=0.9\columnwidth]{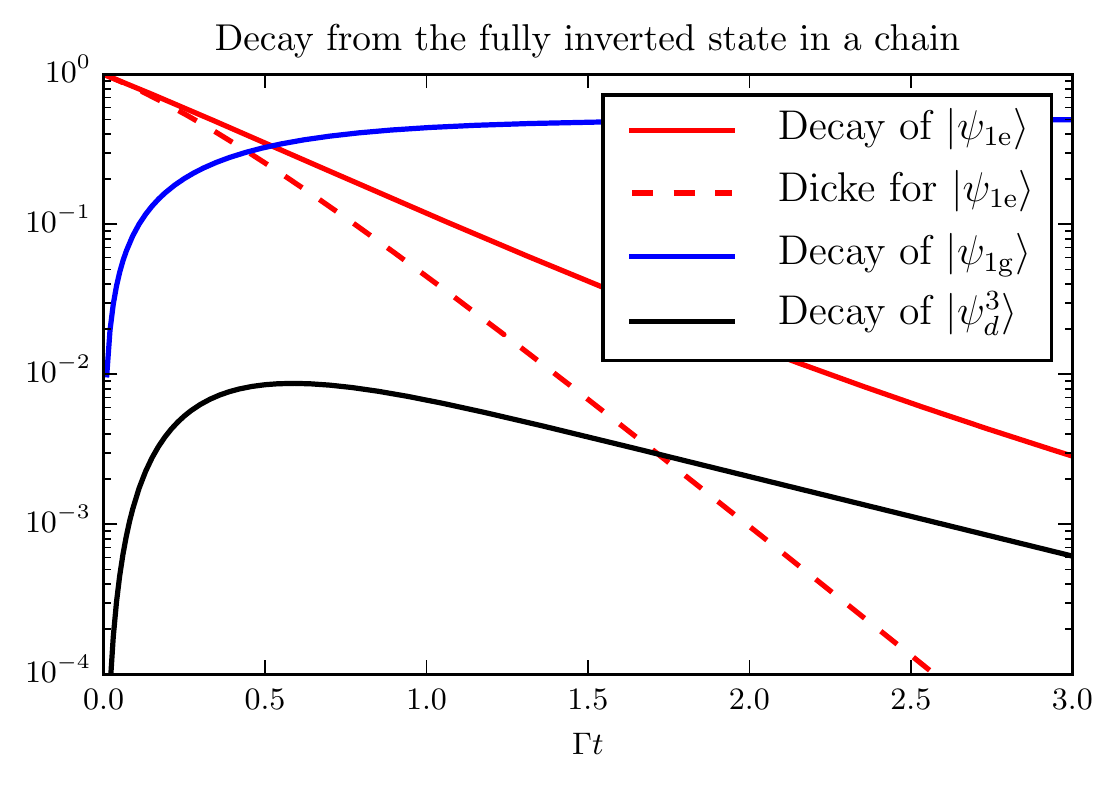}
 \caption{ \emph{ Decay of three $\Lambda$-type atoms in an equidistant chain of distance $\approx \lambda/4$ with nonequal $\Gamma^{ik}$ starting from the totally inverted state $ \ket{eee}$. The red line gives the excited state population per atom, the blue line shows the population of the two ground states $\left( \ket{g_1}+\ket{g_2} \right)/\sqrt{2}$ and the black line gives the dark state $\ket{\psi_\text{d}^3}$ fraction during the decay. For comparison the red dashed line exhibits ideal collective atomic decay with all equal $\Gamma^{ik}$ (Dicke case). Note, that during the evolution the dark (gray) state, which decays much slower, becomes populated partially.} }
 \label{fig5:chaininverted}
\end{figure}

A central question in our configuration concerns the extend to which a dissipative preparation works as well for the case of several independent decay channels. Let us point out that, by construction, it is obvious that the dark state, $\ket{\psi_\text{d}^N}$, is also decoupled from any further symmetric laser excitation, i.e. an excitation by a laser on any transitions using an equal phase on each atom. Hence the state is also dark in absorption measurements similar to coherent population trapping in the ground state manifold~\cite{dalton1985coherent}.

In a first approach to populating the dark state one can simply start from the totally inverted state $\ket{eee}$ for three atoms placed at a suitable finite distance, where the off diagonal elements of the matrix $\Gamma^{ik}$ acquire negative values. In Fig.~\ref{fig5:chaininverted} we demonstrate this mechanism for a three qutrit chain with a distance of about $\lambda/4$. A comparison of the excited state fraction decay for the finite sized chain (red line) with the ideal collective decay (dashed red line) shows  a slowdown of the decay at later times, where a small fraction of the population ends up in the dark state (black line). As the dark state has only a very small overlap with any product state, this fraction is small but can become relatively important at late times. Since the ideal dark state $\ket{\psi_\text{d}^3}$ acquires a finite lifetime at finite distances, this fraction decays as well but at a much slower rate.

\textbf{\emph {Conclusions:}} As our key result we show that the concept of dark or subradiant states can be generalized to multiple decay channels, if one includes one more particle than decay channels. The corresponding dark states are completely anti-symmetric, highly entangled multi-partite states with a plethora of quantum information applications. They can be prepared by a sequence of a bipartite or tripartite gates or via tailored spontaneous emission from multiply excited states in optical lattices. A generalization to multiple excitations and several excited states can be envisaged.

\textbf{\emph {Acknowledgements}} We thank M. Moreno-Cardoner, D.\ Chang and R. Kaiser for helpful discussions and acknowledge
support from DARPA (LO and HR) and the Austrian Science Fund (BK and MH)  (FWF) grants  Y535-N16 and DK-ALM: W1259-N27.

\bibliography{citations}

\section{appendix} 
\label{Appendix_I}
\textbf{\emph {Proofs of some properties of $\ket{\psi_\text{d}^N}$:}}
In this appendix, we prove some of the statements about the state $\ket{\psi_\text{d}^N}$, given in Eq.~(\ref{psidarkn}), made in the main text.

First, we show, that the state $\ket{\psi_\text{d}^N}$ is the unique state (up to superpositions with states containing no excitations) that is stationary for a system consisting of $N$ atoms with $N-1$ independent optical transitions between the upper state $\ket{e}$ and $N-1$ lower states $\ket{g_i}$. To this end, we make use of the criterion for pure states to be stationary under $\mathcal{L}$ derived in~\cite{kraus2008stationaryconditions}. These conditions read as follows.
\begin{enumerate}
	\item  $Q^\dagger \ket{\Phi} = \lambda \ket{\Phi}$ for some $\lambda \in \mathbb{C}$,
	\item $c_l = \lambda_l  \ket{\Phi}$ for some $\lambda_l \in \mathbb{C}$ with $\sum_{l} g_l \left| \lambda_l \right|^2 = \operatorname{Re}(\lambda)$,
\end{enumerate}
where $Q = P - i H$, $P = \sum_l g_l c_l^\dagger c_l$, and $c_l$ are the quantum jump operators with dissipative rates $g_l$~\cite{kraus2008stationaryconditions}. Here, we have that $\{c_l\}_l = \{ S_l^-\}_l$, with the jump operators $S_l^- = \sum_j \sigma_{l, j}^{-}$. As the $S_l^-$ are nilpotent we get $\lambda_l = 0 \forall l$. 
Making a general ansatz for $\ket{\Phi}$, $\ket{\Phi} = \sum_{i_1, \ldots, i_N} c_{i_1, \ldots, i_N} \ket{i_1, \ldots, i_N}$, we get the necessary condition $c_{i_1, \ldots, i_N} = 0 \; \forall i_1, \ldots, i_N$, where at least one $i_k = e$ \footnote{Note that states containing no excitations may always be added, as they are stationary} and no $i_k = g_l$, for $\ket{\Phi}$ to be an eigenstate of $S_l^-$. As, according to the conditions, this must hold for all $l$, we get $c_{i_1, \ldots, i_N} = 0$, unless $\{i_1, \ldots, i_N\}=\{e, g_1, \ldots, g_{N-1}\}$. Hence, $\ket{\Phi} = \sum_{\pi \in S_n} \tilde{c}_{\pi} \ket{\pi(1), \ldots, \pi(N)}$. Furthermore, by applying $S^-_l$ onto $\ket{\Phi}$ one can derive the necessary conditions 

\begin{align}
c_{i_1, \ldots, i_{r-1}, e, i_{r+1}, \ldots i_{s-1}, g_l, i_{s+1}} + \\ \nonumber + c_{i_1, \ldots, i_{r-1}, g_l, i_{r+1}, \ldots i_{s-1}, e, i_{s+1}}  = 0 \; \forall i_j.
\end{align}

 Using this argument consecutively three times, it follows that $\tilde{c}_{\pi} = - \tilde{c}_{\sigma \circ \pi}$, for arbitrary transpositions $\sigma$. This implies $\ket{\Phi} \propto \ket{\psi_\text{d}^N}$ and hence completes the proof.

Let us now show that the geometric measure of entanglement of $\ket{\psi_\text{d}^N}$ is given by $E_g(\ket{\psi_\text{d}^N}) =1- \max_{\ket{a_1}, \ldots, \ket{a_N}} \left|\braket{a_1, \ldots, a_N|\psi_\text{d}^N} \right|^2 = 1- \frac{1}{N!}$ (see also~\cite{hayashi2008geometric}). To this end, we explicitly find one of the product states $\ket{a_1, \ldots, a_N}$ that maximize the overlap $\left|\braket{a_1, \ldots, a_N|\psi_\text{d}^N} \right|$. In order to find such a state, note that due to the symmetry $S^{\otimes N}$ of $\ket{\psi_\text{d}^N}$, we can choose $\ket{a_N}$ arbitrarily and still get maximal overlap $\left|\braket{a_1, \ldots, a_N|\psi_\text{d}^N} \right|$ by choosing $\{\ket{a_1}, \ldots, \ket{a_{N-1}}\}$ optimal. Let us choose $\ket{a_N} = \ket{N-1}$. We obtain $\braket{a_N|\psi_\text{d}^N} = \frac{1}{N} \ket{\psi_\text{d}^{N-1}}$. In order to maximize the overlap $\frac{1}{N} \left|\braket{a_1, \ldots, a_{N-1}|\psi_\text{d}^{N-1}} \right|$, we subsequently have to choose $\{\ket{a_i}\}_{i \in {1, \ldots, N-1}}$ in $\operatorname{span}\{\ket{0}, \ldots, \ket{N-2}\}$. The procedure is repeated recursively in order to fix $\ket{a_1, \ldots, a_N} = \ket{0, \ldots, N-1}$ and obtain $E_g(\ket{\psi_\text{d}^N}) = 1-\frac{1}{N!}$.

\section{appendix}
\label{appendix2}
\begin{figure}
 \includegraphics[width=0.85\columnwidth]{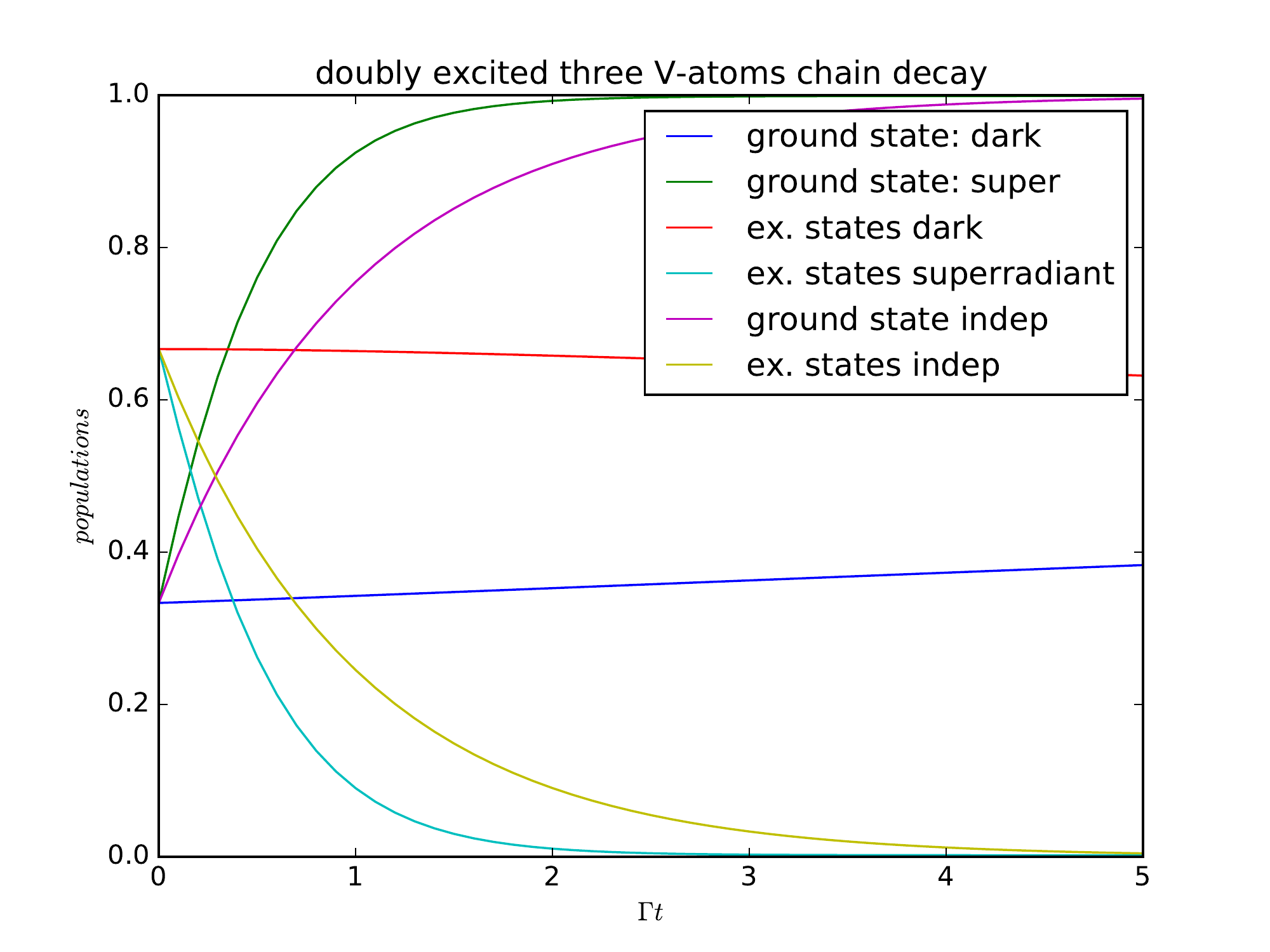}
 \caption{\emph{Decay of three V-type atoms in small chain $d \ll \lambda$ with initially two excitations. Blue gives the ground state and red and the excited state populations. The other lines show the dynamics for a super-radiant initial state for comparison.} }
 \label{fig:chainv-system}
\end{figure}

\textbf{\emph {Generalization to V-systems: two excitation dark state}}
A second generic configuration of a three level system with decay is a V type system, where two upper states $(\ket{e_1},\ket{e_2})$ decay to the same ground state $\ket{g}$. This is typically the case for a zero angular momentum ground state. Interestingly it turns out that the dark state $\ket{\psi_{de}^3}$ constructed in the same way as above by replacing the states g and e is also a non decaying states. In contrast to the case investigated in this manuscript it, however, stores two excitation quanta in the three atoms. This is shown in Fig.~\ref{fig:chainv-system}, where we compare its slow decay (red line) with independent atoms (yellow) and super-radiant decay (light blue). It is an additional interesting question, whether this could be generalized to storing $N-1$ excitations in $N$ atoms creating an almost inverted collective dark state in a realistic setup.  

\end{document}